\documentstyle[prl,aps]{revtex}
\begin{document}
\twocolumn[\hsize\textwidth\columnwidth\hsize\csname@twocolumnfalse\endcsname
\author{A.E. Koshelev}
\address{Argonne National Laboratory,Argonne, IL 60439,
and\\ 
Institute of Solid State Physics, Chernogolovka, Moscow District, 142432, Russia}
\title{Mechanism of thermally activated c-axis dissipation in layered High-T$_c$
superconductors at high fields}
\maketitle
\begin{abstract}
We propose a simple model which explains experimental behavior of $c$-axis
resistivity in layered High-T$_c$ superconductors at high fields in a limited temperature range.  It is
generally accepted that the in-plane dissipation at low temperatures is caused
by small concentration of mobile pancake vortices whose diffusive motion is
thermally activated. We demonstrate that in such situation a finite conductivity
appears also in $c$-direction due to the phase slips between the planes caused
by the mobile pancakes.  The model gives universal relation between the components of
conductivity which is in good agreement with experimental data. 
\end{abstract}
\twocolumn 
\vskip.2pc] 
\narrowtext 

A well known property of layered High-T$_c$ superconductors (e.g., 
$Bi_2SrCa_2Cu_2O_x$ (BiSCCO)) is an existence of extended 
low-temperature tails in the temperature dependencies of the both 
in-plane \cite{rho_ab,Busch} and out-of-plane 
\cite{Busch,Briceno,Latyshev,Cho} components of resistivity for 
magnetic field applied along the $c$-axis.  An important experimental 
fact is that the thermally activated behavior for both $\rho _{ab}$ 
and $\rho _c$ is characterized by the same value of activation energy 
(see Ref.~\onlinecite{Latyshev} and Fig.~\ref{FigRho_c_ab}).  This 
indicates that dissipations in both directions have {\it essentially 
the same origin}.

The origin of the tail for the in-plane resistivity component is 
qualitatively understood.  It is theoretically expected that at fields 
larger than the crossover field $B_{cr}\approx \Phi_0/(\gamma s)^2$ 
and temperatures larger than the decoupling temperature $T_{dec}$ the 
system can be treated as a set of almost independent two-dimensional 
(2D) vortex lattices in the layers \cite{GlazKosh} (here $\gamma$ is 
the anisotropy of the London penetration depth and $s$ is the 
interlayer spacing).  The in-plane dissipation can therefore be 
attributed to the thermally activated motion of point defects in the 
2D lattices (weak disorder) or weakly pinned mobile pancake vortices 
(strong disorder) \cite{TAFF}.  For a 2D superconductor with weak 
pinning a crossover between free flux flow and thermally activated 
motion should take place at the melting temperature of the vortex 
lattice $T_m \approx 0.014s\Phi_0^2/(4\pi\lambda)^2$ as observed in 
the thin films of $a-Nb_3Ge$ \cite{MeltNbGe} and in numerical 
simulations \cite{MySym}.  This seems not to be the case for 
available samples of the  
BiSCCO compound where crossover to the thermally activated behavior at 
high fields ($>$ 1 T) takes place at temperatures ~ 50-70 K which is 
significantly larger than the estimate for the melting temperature 
$T_m\approx 13 K$ obtained using experimental values of the London 
penetration depth $\lambda_{ab}\approx 1800\AA$ and $s=15\AA$.  This 
indicates that in available BiSCCO single crystals pinning is strong 
enough to destroy the melting transition in the field range ~1-10 T 
and the crossover temperature between free flux flow and thermally 
activated regime (~50-70 K) has to be interpreted as the depinning 
temperature of a single pancake vortex renormalized by the intervortex 
interactions \cite{LowFields}.  In this case a very natural assumption 
is that the lattice is strongly pinned and the in-plane dissipation is 
caused by the motion of mobile pancake vortices with the small 
concentration $n$ and diffusion constant $D$.  In-plane conductivity 
within this model can be estimated as follows.  In-plane transport 
current $j$ drives mobile pancakes with velocity $v=\mu s 
(\Phi_0/c)j$.  Here $\mu$ is the average pancake mobility, which is 
connected with the diffusion constant $D$ by the Einstein relation, 
$D=\mu T$.  Pancake motion produces electric field $E=(\Phi_0/c)vn$.  
Therefore the conductivity in {\it ab}-plane $\sigma _{ab}$ in this model is 
given by
\begin{equation}
\label{Sig-ab}
\sigma _{ab}=\left( \frac c{\Phi _o}\right) ^2\frac{T}{snD}
\end{equation}
Both $n$ and $D$ presumably have Arrhenius-type temperature 
dependencies but their temperature dependence as well as detailed 
mechanism of in-plane motion have no influence for our further 
consideration.

In the region where the conductivity in the $c$-direction is much larger than
the normal conductivity it has the Josephson origin and is caused by phase slips
between neighboring layers.  It was assumed \cite{Briceno,Gray} that these phase
slips are similar to the phase slips in a single Josephson junction with the
typical area $\Phi_0/B$ and therefore the Ambegaokar-Halperin formula for the
resistivity of small Josephson junction can be used to describe the temperature
and field dependencies of $\rho_c$. Unfortunately this physically transparent
assumption does not have microscopic support.  

Diffusive motion of pancake vortices at high enough temperatures leads 
to the time variations of the phase difference between the layers.  
Such pancake-induced phase slips provides a candidate mechanism for 
the c-axis dissipation which was not considered so far.  In this 
Letter we obtain c-axis conductivity due to such mechanism.  The key 
assumption of our model is that the phase slips are caused the 
diffusive motion of a small amount of pancake vortices independently 
in different layers.  The model is applicable above the decoupling 
temperature where the global superconducting coherence is broken.  
This means that the model can be used to describe dissipation in a 
limited temperature range ($\sim$ 40-70 K).

The starting point of our calculation is the Kubo formula which 
relates $\sigma _c$ with the correlation function of the Josephson 
current:
\begin{equation}
\label{Sig-c}
\sigma _c=\frac{sj_J^2}T\int drdt\left\langle \sin \delta \phi
(0,0)\sin \delta \phi ({\bf r},t)\right\rangle
\end{equation}
Here $j_J$ is the Josephson current, $\delta 
\phi=\phi_2-\phi_1-\frac{2\pi s}{\Phi_0}A_z$ is the Gauge invariant 
phase difference between neighboring layers; the phases 
$\phi_{1,2}({\bf r},t)$ are mainly determined by the vortex positions 
${\bf R}_{1,2i}$ in the layers, $\phi_{1,2}({\bf 
r},t)=\sum_i\phi_v({\bf r}-{\bf R}_{1,2i}(t))$ where $\phi_v({\bf 
r})={\rm atan}(y/x)$ is the phase distribution around the core of a 
single vortex.  We calculate $\sigma_c$ in the lowest order with 
respect to the interplane Josephson coupling. Within this 
approximation one can neglect correlations between Josephson currents 
in different layers, which are already excluded in Eq.~(\ref{Sig-c}).  
We can also neglect interplane phase correlations in the sine-sine 
correlation function in Eq.~(\ref{Sig-c}).  This means that averages 
like $\langle \cos \delta \phi \rangle$ assumed to be zero and
\begin{eqnarray}
\nonumber \langle \sin \delta \phi (0,0)\sin \delta \phi ({\bf 
r},t)\rangle\approx(1/2)\hbox{Re}[S({\bf r},t)],\\
\nonumber \hbox{with } S({\bf 
r},t)=\left\langle \exp i\left(\delta \phi ({\bf r},t)-\delta\phi 
(0,0)\right)\right\rangle.  
\end{eqnarray}
As follows from Eq.~(\ref{Sig-c}) the value of the out-of-plane 
conductivity is determined by how fast the phase correlations decrease 
in space and time.  The origins of decay of the phase correlations in 
space and time are very different.  The decay in space is determined 
by the randomness in the vortex arrangement while the decay in time is 
determine by the vortex motion.  In our model the static phase 
difference $\delta \phi (0,0)-\delta
\phi ({\bf r},0) $ is established by the almost all rigidly 
pinned vortices \cite{DID} while dynamic behavior is caused by the 
small amount of mobile vortices.  In such a situation the correlation 
function can be splitted into the separately averaging static and 
dynamic parts, $S({\bf r},t)\approx S({\bf r})S(t)$ and 
Eq.~(\ref{Sig-c}) becomes,
\begin{equation}
\label{Sig-cAppr}
\sigma _c\approx \frac{sj_J^2}{2T}\int drdt S({\bf r})S(t) 
\end{equation}
with 
\begin{eqnarray}
\nonumber S({\bf r})&=&\left\langle \exp \left[ i\left( \delta \phi 
(0,0)-\delta \phi ({\bf r},0)\right) \right] \right\rangle,\\
\nonumber S(t)&=&\left\langle \exp \left[ i\left( \delta \phi ({\bf 0}
,0)-\delta \phi ({\bf 0},t)\right) \right] \right\rangle. 
\end{eqnarray}
In absence of correlations between vortex positions in neighbor layers 
the static part is expected to decay at distances of the order of the 
average intervortex spacing $a_0$.  This assumption can be supported 
by direct calculation for the case of randomly placed pancakes, which 
gives \cite{DID}
\begin{equation}
\label{Static}
S({\bf r})\approx \exp \left( -\pi
n_vr^2\ln(r_{max}/r)\right) 
\end{equation}
with $n_v=B/\Phi_0$.  We assume that the phase at given point is not 
sensitive to pancake displacements at distances larger than some phase 
coherence distance $r_{max}$, which provides upper cutoff to the 
logarithmically diverging integrals.  The static phase correlation 
function for randomly placed pancakes Eq.~(\ref{Static}) actually has 
the fastest possible descend.  Correlations in the pancake arrangements 
always increase the spatial range of $S({\bf r})$.  Decay of the dynamic 
part is determined by the diffusion of mobile pancakes and is given by
\begin{equation}
\label{Dynamic}
S(t) =\exp \left(-2\pi nDt\ln(r_{max}^2/Dt)\right)
\end{equation}
Substituting Eqs.~(\ref {Static}) and (\ref{Dynamic}) into 
Eq.(\ref{Sig-cAppr}) we obtain \cite{Validity}
\begin{equation}
\label{Sig_cRes}
\sigma _c\approx 
\frac1{2\pi \ln(r_{max}/a_0)\ln (nr_{max}^2)}\frac{sj_J^2}{n_vTnD} 
\end{equation}
Up to logarithmic factors and numerical constant this result immediately 
follows from Eq.~(\ref{Sig-c}) assuming that the interlayer phase 
difference is correlated at distances of the order of the average vortex 
spacing and at times of the order of the typical ``phase slip'' time 
$t_{ps}=1/nD$.
Using the relation $j_J=\frac{2\pi c}{\Phi _o}E_J$ and Eq.  
(\ref{Sig-ab}) the last equation can be transformed to
\begin{equation}
\label{Ratio} 
\sigma _c\approx C\frac{\Phi_0s^2E_J^2}{BT^2}\sigma _{ab}
\end{equation}
with $C=\frac{2\pi }{\ln(nr_{max}^2)\ln(r_{max}/a_0)}$.  
Therefore within our model the ratio of conductivities depends only 
upon material parameters, temperature, and field, and does not depend 
upon the details of the mechanism, responsible for dissipation.  
Eq.~(\ref{Ratio}) represents the main result of our Letter.

The model can be verified independently by measuring the frequency 
dependence of $\sigma_c$.  The frequency dependence can be taken into 
account by adding the factor $\cos (\omega t)$ under the integral in 
Eq.(\ref{Sig-c}).  This gives Lorentzian frequency dependence $\sigma_c(\omega)$:
\begin{equation}
\label{omega}
\frac{\sigma_c(\omega)}{\sigma_c(0)}=\frac{\omega_{ps}^2}{\omega^2+\omega_{ps}^2}
\end{equation}
The typical ``phase slip'' frequency $\omega_{ps}=2\pi 
nD\ln(nr_{max}^2)$ can be directly related to the in-plane component 
of resistivity $\rho_{ab}=1/\sigma_{ab}$ using Eq.~(\ref{Sig-ab}), 
which gives $\omega_{ps}\approx 2\:10^8 [1/{\rm s}]\cdot T[{\rm 
K}]\cdot
\rho_{ab}[\mu
\Omega \cdot {\rm cm}]$. The direct correlation between typical 
frequency in $\sigma_c(\omega)$ and $\rho_{ab}$ provides additional  
experimental check of the theory.

Let us discuss the limitations of our model. At low enough 
temperature slowing down of the pancake motion in the layers leads to 
increase of the interlayer correlations so that the diffusion of 
pancakes in different layers can not be considered independent any 
more. Quantitatively the influence of the Josephson coupling on the 
diffusive motion is characterized by value of the Josephson force acting on 
pancake. This force is determined by the variation of Josephson energy 
under small pancake displacement and can be calculated as
\begin{equation}
\label{fJ}
f_{J\alpha }({\bf R},t)=-E_J\int d^2r\frac{\partial \phi _v({\bf r}-{\bf 
R}(t))}{\partial r_\alpha }\sin \delta \phi ({\bf r},t)
\end{equation}
Neglecting again the interplane phase correlation we obtain the correlation 
function of this force
\begin{equation}
\label{fJcorr}
\left\langle f_{J\alpha }(0)f_{J\alpha ^{\prime }}(R)\right\rangle
\approx \delta _{\alpha \alpha ^{\prime }}
\frac{\pi E_J^2}{n_v}\frac{\ln (r_{\max }/R)}{\ln (r_{\max }/a)} S(t)
\end{equation}
As one can see from this expression the Josephson force has typical 
amplitude $f_J\sim aE_J$ and is correlated at times $\sim t_{ps}$ 
and at large distances of the order of $r_{max}$.  This means that 
mobile pancakes within the correlation area $\sim r_{max}^2$ will 
drift under this force with approximately the same velocity $v\approx 
\mu f_J$ during the ``phase slip'' time $t_{ps}$.  
This drift motion will produce an extra phase change $\delta \phi 
\approx nr_{max} \mu f_J t_{ps}\approx r_{max}f_J/T$.  One can neglect 
the Josephson coupling and treat pancake motion in different layers 
independently when this extra phase shift is small, which gives the 
condition $r_{max}f_J\ll T$.  More precise estimate requires 
evaluation of the in-plane phase coherence distance $r_{max}$. This 
is beyond the scope of the simple model proposed in this Letter.

Several mechanisms limit applicability of this model 
at high temperatures.  Quasistatic phase fluctuations with the 
amplitude $\chi =\left\langle (\delta \phi )^2\right\rangle/2$ lead to 
additional suppression of the Josephson energy, which can be described 
by the Debye-Waller factor $E_J^{eff}\approx E_J\exp \left( - \chi 
\right)$, and reduces the ratio $\sigma _c/\sigma _{ab}$ as compared 
to Eq.(\ref{Ratio}).  Two contributions to $\chi$ come from thermal 
oscillations of the pinned pancakes ($\chi_1$) and from the spontaneous 
interplane phase fluctuations not related to pancake motion ($\chi 
_2$), $\chi = \chi _1+ \chi _2$. We estimate this contributions as
$$
\chi_1\approx\frac{ n_vT}{\alpha _L}\ln \frac{r_{\max }^2\alpha _L}{T};
\quad \chi_2=\frac {T}{4J} \ln\frac{J}{E_J^{eff}\xi^2}
$$ where $\alpha _L\approx s\Phi_0 j_c/(c\xi)$ is the Labush 
parameter, $j_c$ is the critical current, and $J=s\Phi _0^2/(\pi (4\pi 
\lambda )^2)$ is the in-plane phase stiffness.  Taking 
$j_c=10^6$A/cm$^2$, $T=60$K, $B=1$T, $\ln (r_{\max }^2\alpha _L/T)= 5$ 
we obtain the estimate $\chi _1\approx 0.15$.  Using parameters of 
BiSCCO and taking the logarithmic factor in $\chi _2$ as 5 we obtain 
$\chi _2\approx 4[K]/(T_c-T)$.  Therefore the spontaneous interplane 
phase fluctuations can be ignored at temperatures $\sim$ 10 K below 
$T_c$.

Another important factor, which can not be ignored at high 
temperatures, is the quasiparticle conductivity.  This conductivity 
has to be added to the phase slip conductivity (\ref{Sig-c}) and 
therefore it enlarges the conductivity ratio in comparison with prediction 
of Eq.(\ref{Ratio}).

To verify the prediction of the model we plot in Fig.~\ref{FigRatio} 
temperature dependencies of the conductivity ratio at different fields 
(the data \cite{Busch} are the same as in Fig.~ (\ref{FigRho_c_ab})).  
One can see that within the temperature range 50-70 K for fields 0.5-2 
T a general trend is in agreement with Eq.~(\ref{Ratio}), i.e.  the 
ratio decreases with temperature and field.  Taking the conductivity 
anisotropy at 60 K and 1 T ($\approx 2.7\cdot 10^{-6}$) and using the 
relation between the Josephson coupling energy and the Ginzburg-Landau 
anisotropy ratio $\gamma$, $E_{\rm J}=\Phi_0^2/\pi (4\pi \lambda 
\gamma)^2s$, we estimate from Eq.  (\ref{Ratio}) the value of $\gamma$ 
as 370 which is within the range reported in the literature 
\cite{gamma}.  An abrupt increase of the ratio at low temperatures 
indicates probably the emergence of the global coherence in the system 
and the glass transition. 

To verify quantatively the validity of Eq.~(\ref{Ratio}) we need to 
know the temperature dependence of the Josephson coupling energy.  
This dependence is determined by the temperature dependencies of the 
tunneling integral $t$ and superconducting order parameter $\Psi$, 
$E_J\propto t|\Psi|^2$.  Usually the first dependence is neglected.  
However, as recently was proposed by Bulaevskii {\it et al}
\cite{rho_cT}, the semiconducting temperature dependence of the 
normal-state c-axis conductivity $\sigma_{cn}$ at low temperatures can 
be ascribed to the temperature dependence of $t$.  We therefore assume 
that $t$ and $\sigma_{cn}$ have the same temperature dependence.  The 
temperature dependence of $\sigma_{cn}$ have been recently measured by 
Cho {\it et al} \cite{Cho} for the temperatures down to $\approx$ 55 K 
using high magnetic fields (up to 18 T).  This dependence in the 
interval 55-90 K can be very well fitted by the exponential formula 
$\sigma_{cn}\propto \exp(T/29[K])$.  Therefore to verify 
Eq.~(\ref{Ratio}) we try to fit the conductivity ratios by the formula
\begin{equation}
\label{Fit}
\sigma_c/\sigma_{ab}(B,T)=A(B)\exp(2T/29)(1-T/T_c)^2T^{-2}
\end{equation}
with $A(B)$ being the fitting parameter. The factor $1-T/T_c$ accounts for
temperature dependence of $|\Psi|^2$.  As one can see this formula gives very
good fit to the data in the temperature range 45-70 K. However the field
dependence of the fitting parameter $A(B)$ is weaker than predicted by
Eq.~(\ref{Ratio}), $A(B)\propto B^{-0.52}$.  The most probable origin of this
discrepancy is that increasing of the field reduces randomness in vortex
arrangement within the layers and, as a consequence, the static phase
correlations decay in space slower than is predicted by Eq.~(\ref{Static}). 

In conclusion, we propose simple model, which provides the mechanism for thermally
activated dissipation along the c-axis in the vortex state of disordered layered 
superconductors. The model gives the universal relation between the conductivity 
components.

I would like to thank A.I.Larkin for useful discussion.
This work was supported by the National Science Foundation Office of the 
Science and Technology Center under contract No.  DMR-91-20000. and by the 
U. S. Department of Energy, BES-Materials Sciences, under contract No. W-31-
109-ENG-38.

\newpage
\centerline{\bf FIGURE CAPTIONS} 
\begin{figure}
\caption{Temperature dependencies of the resistivity components $\rho_{ab}$ 
and $\rho_c$ for BiSCCO for fields 0.5,1,2,and 5 T.  Data are taken 
from Ref.~\protect\onlinecite{Busch}.  Note that in spite of huge 
difference in absolute values, $\rho_{ab}$ and $\rho_c$ have almost 
identical temperature dependencies.}
\label{FigRho_c_ab}
\end{figure}
\begin{figure}
\caption{Temperature dependencies of the conductivity ratio for the data from 
Ref.~\protect\onlinecite{Busch}.  Solid lines give the fits by the 
formula (\protect\ref{Fit}), which were made to verify the prediction of 
Eq.(\protect\ref{Ratio}). Table gives the values of parameter $A$ at 
different fields obtained from the fits. }
\label{FigRatio}
\end{figure}


\begin{references}
\bibitem{rho_ab}T.T.M. Palstra {\it et al}, Phys.Rev.Lett. {\bf 61}, 1662 (1988);
N.Kobayashi {\it et al}, Physica C{\bf 159},295 (1989) 
\bibitem{Busch} R.Busch{\it et al}, Phys.Rev.Lett. {\bf 69}, 522 (1992)
\bibitem{Briceno}G.Brice\~no, M.F.Crommie, and A.Zettl, 
Phys.Rev.Lett. {\bf 66}, 2164 (1991)
\bibitem{Latyshev}Yu.I.Latyshev and A.F.Volkov, Physica C{\bf 182},47 (1991)
\bibitem{Cho}J.H.Cho {\it et al},Phys. Rev. B, {\bf 50},6493 (1994)
\bibitem{GlazKosh} L.I.Glazman and A.E.Koshelev, Phys. Rev. B, {\bf 43},2835
(1991)
\bibitem{TAFF}P.H.Kes {\it et al}
Supercond. Sci. Technol.  {\bf 1}, 242 (1989)
\bibitem{Gray} K.E.Gray and D.H.Kim Phys.Rev.Lett. {\bf 70}, 1693 (1993)
\bibitem{MeltNbGe}P.Berghuis {\it et al}, Phys. Rev.  Lett.  {\bf 65}, 2583,
(1990); P.Berghuis, and P.H.Kes, Phys.Rev.  B {\bf 47}, 262,(1993)
\bibitem{MySym}A.E.Koshelev, Physica C {\bf 198}, 371 (1992) and to be published
\bibitem{DID}A.E.Koshelev, L.I.Glazman, and A.I.Larkin Phys. Rev. B, 
{\bf 53},????, (1996)
\bibitem{LowFields} We should emphasize that this statement is related to large
fields where the melting temperature is close to the melting point of a single
layer. As field decreases the melting temperature increases due to 3D effects
and at some point becomes larger than the depinning temperature. This means that
the melting transition is expected to be restored at small fields ($<$ 1 kOe). 
Ordered vortex lattice at small fields indeed was observed by decorations
(R.N.Kleiman Phys.Rev.Lett. {\bf 62}, 2331 (1989)), neutron diffraction 
(R.Cubitt {\it et al} Nature(London) {\bf 365}, 407 (1993)), 
and Lorentz Microscopy (Harada {\it et al} Phys.Rev.Lett.  {\bf 71}, 
3371 (1993))
\bibitem{Validity}Note that Eqs.~(\protect\ref{Static}) and 
(\protect\ref{Dynamic}) give $\langle \exp i \delta \phi \rangle=0$, 
which is valid only if the interlayer coupling is completely 
neglected.  If small interlayer coupling is included than the 
correlation function $S({\bf r},t)$ does not vanish at large distances 
and times but saturates at $S({\bf \infty },\infty )=\langle \cos 
\delta \phi \rangle^2\ll1 $.  This large-scale behavior does not 
influence much the integral in Eq.(\protect\ref{Sig-c}) and result 
(\protect\ref{Sig_cRes}), because in higher order with respect to 
interlayer coupling one has to keep term $\left\langle \exp i 
[\delta\phi(0,0)+\delta \phi ({\bf r},t) ]\right \rangle$, which was 
skipped in the transformation from Eq.(\protect\ref{Sig-c}) to 
Eq.(\protect\ref{Sig-cAppr}).  This term term being added in 
Eq.(\protect\ref{Sig-cAppr}) exactly cancels $S({\bf \infty },\infty 
)$ at large scales and the main contribution still comes from small 
distances and times where approximations (\protect\ref{Static}) and 
(\protect\ref{Dynamic}) are applicable.
\bibitem{gamma}K.Okuda {\it et al}, J.Phys.Soc.Jpn., {\bf 60}, 3226, (1991);
J.Martinez {\it et al}, Phys.Rev.Lett. {\bf 69}, 2276 (1992); F.Steinmeyer
{\it et al}, Physica C{\bf 194-196},2401 (1994)
\bibitem{rho_cT}L.N.Bulaevskii {\it et al}, unpublished

\end{references}
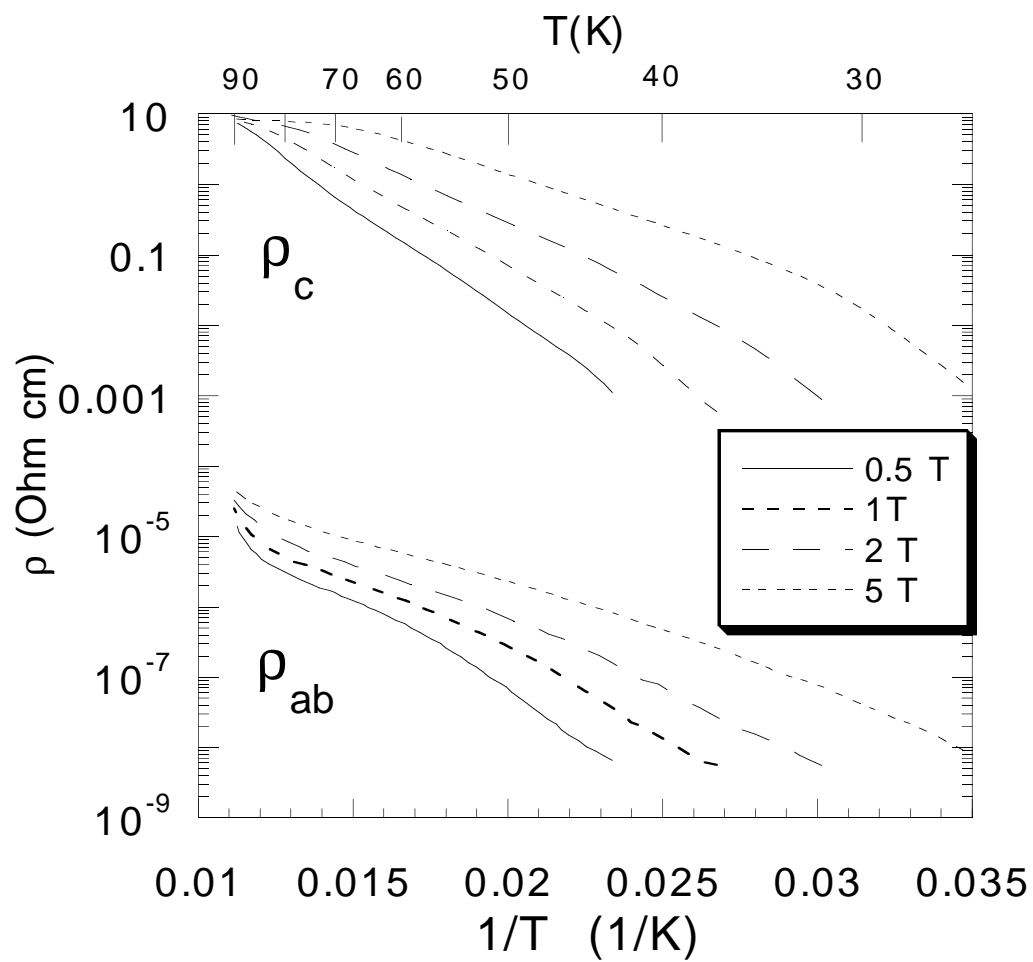
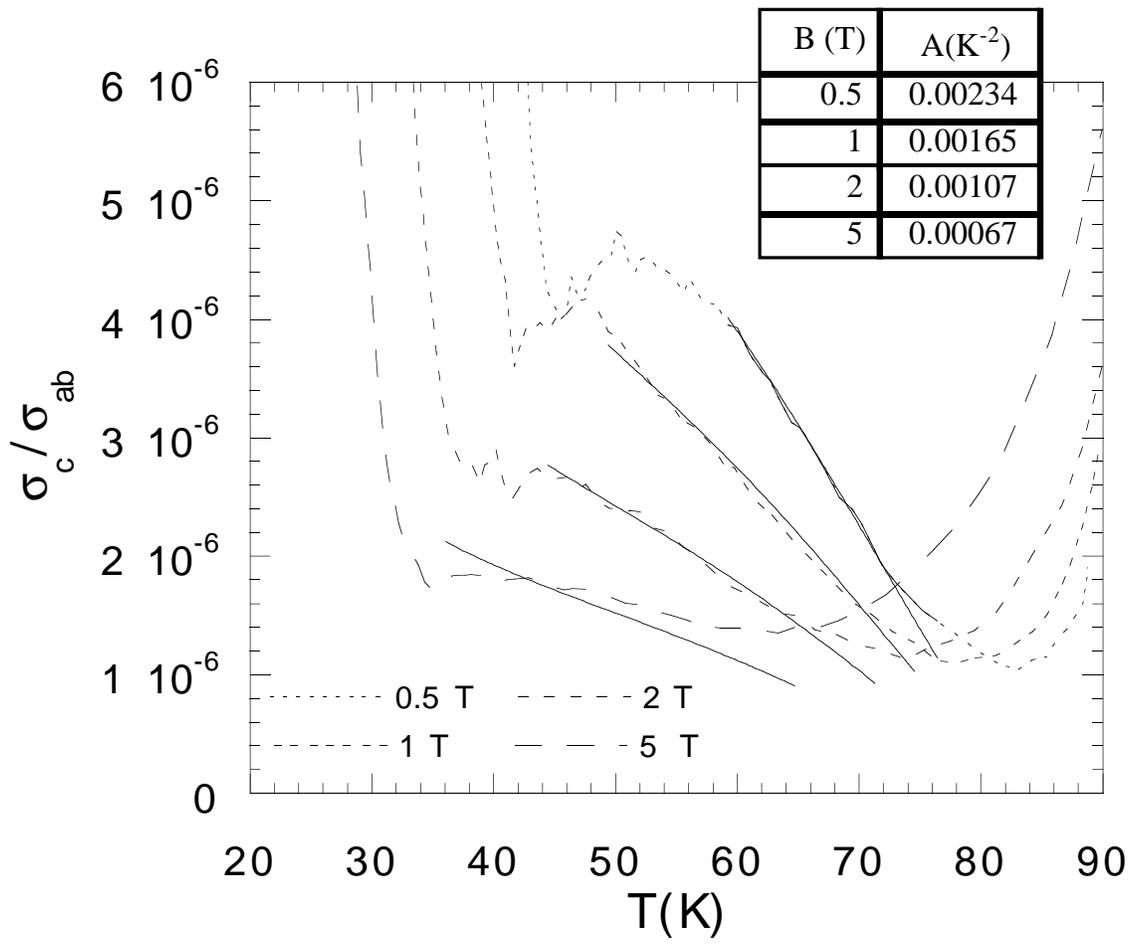
\end{document}